\documentclass[a4paper,11pt] {article}
\usepackage[latin9]{inputenc}
\usepackage{color}
\usepackage{amsmath}
\usepackage{amssymb}
\usepackage{caption}
\usepackage{subcaption}
\usepackage{graphicx}
\usepackage[authoryear]{natbib}
\usepackage[markup=underlined]{changes}
\usepackage[unicode=true,
 bookmarks=false,
 breaklinks=false,pdfborder={0 0 1},backref=section,colorlinks=true]
 {hyperref}
\hypersetup{
 linkcolor=blue,citecolor=blue}

\makeatletter

\date{}

\usepackage{setspace}
\usepackage{lscape}
\usepackage{multicol}
\usepackage{amsthm}
\usepackage{bm}
\usepackage[toc]{appendix}
\numberwithin{equation}{section}
\usepackage{eucal}
\usepackage{lscape}
\usepackage{comment}
\usepackage{float}
\usepackage{epstopdf}
\usepackage{booktabs}
\usepackage[export]{adjustbox}
\usepackage[hmargin=2.5cm,vmargin=3.5cm]{geometry} 
\usepackage[normalem]{ulem}
\renewcommand{\baselinestretch}{1.0} 
\makeatother

\begin{document}
\title{\vspace{-3.0cm}\textbf{\Large{}
 Spillover Effects of US Monetary Policy on Emerging Markets Amidst Uncertainty}\thanks{\emph{Disclaimer}: The views expressed in this article are those of the authors and do not necessarily represent those of the IMF, its Executive Board, or its management. Lastauskas is grateful for the support from Iceland, Liechtenstein and Norway through the EEA grants (project no. S-BMT-21-8 (LT08-2-LMT-K-01-073)) under a grant agreement with the Research Council of Lithuania (LMTLT). }} 
\normalsize{\author{Povilas Lastauskas\thanks{Department of Business Analytics and Applied Economics, Francis Bancroft Building, Mile End Campus, London, UK; Faculty of Economics and Business Administration, Vilnius University, Lithuania, and Homerton College, University of Cambridge, UK. \ \textit{Email}: P.Lastauskas@trinity.cantab.net; \ \textit{Web}: www.lastauskas.com.}\\
 Queen Mary University of London, Vilnius University, and Homerton College (Cambridge) \and Anh Dinh Minh Nguyen\thanks{International Monetary Fund, Washington, DC 20431, USA. \ \textit{Email}: anguyen3@imf.org; \ \textit{Web}: https://sites.google.com/site/anecon14/home.}\\
International Monetary Fund}}

\maketitle


\thispagestyle{empty} 
\begin{abstract}
This paper examines the impact of US monetary policy tightening on emerging markets, distinguishing between direct and indirect spillover effects using the global vector autoregression with stochastic volatility covering 32 countries. The paper shows that an increase in the US interest rate significantly reduces output for emerging markets, leading to larger, more prolonged, and persistent declines. Such an impact is further intensified by global trade integration, causing a sharper yet slightly quicker rebounding output drop. The spillover effects are significantly amplified when US monetary policy tightening is accompanied by an increase in monetary policy uncertainty. Finally, emerging markets exhibit considerable heterogeneity in their responses to US monetary policy shocks.

\end{abstract}


\setcounter{page}{1}

\newpage
\section{Introduction}

\begin{flushright}
``If the US sneezes, do emerging markets still catch a cold?''  Mark Sobel, Official Monetary and Financial Institutions Forum, August 3 2022.
\par\end{flushright}

\begin{flushright}
``Emerging markets are facing their demons as traders mull whether U.S. Federal Reserve interest rates will rise as high as 6\%, a level that could kick weaker countries when they're down, while diverging global growth paths and China's reopening might cushion some of the blow for the bigger ones.'' Rodrigo Campos, Reuters, 9 March, 2023.
\par\end{flushright}

When the economic clouds are turning darker in the United States (US) and the Fed starts tightening monetary policy to combat the surge in inflation, the attention is often drawn to emerging markets. A rise in the US dollar adds additional complexity for the emerging markets' central banks, grabbling with the local currency's stability vis-a-vis the US dollar. The risk of widening the interest rate differential vis-\`a-vis the US Fed funds translates into more volatile capital flows, asset repricing, thereby creating further pressure on local currency, making local central bank's manoeuvre space more limited, and macroeconomic uncertainty higher. The conventional wisdom often revolves around the ``global financial cycle," a phenomenon that a high degree of co-movement in asset values, capital flows and financial variables is caused by financial centres, in particular, that the US and its monetary policy shocks induce co-movements in the international financial variables (\citealp{Rey2015,Miranda-Agrippino2020,MirandaAgrippino2020,LN:2021}). 

The importance of global drivers and the US economic and financial conditions playing a key role in shaping the global dynamics for emerging markets has been documented extensively in the literature (\citealp{Ahmed2021,CaZorzi2021,Georgiadis2016,Lee2020}). However, the literature is scant in understanding the proportion of the US monetary policy effect attributable to its direct impact versus that transmitted indirectly through other countries also affected by unexpected US interest rate movements.  In other words, emerging markets with even weak direct links to the US can still bear a substantial burden if other economies -- with which they have trade or financial connections -- are impacted by the US and its trade partners. Importantly, when US monetary policy tightens, trade rebalances not only directly but also indirectly. A stronger dollar makes US imports cheaper, but at the same time, many other currencies tend to depreciate. The bilateral trade flows with other economies thus depend on both \textit{direct linkages} as well as \textit{other trade partners} which are exposed to the US economy. Moreover, and  in addition to the real economy adjustments, capital flows also react to US dollar fluctuations as well as risk-adjusted interest rate differentials. If a developing country's central bank does not react or has no action space due to, for instance, high levels of dollar-denominated debt, capital flight to safety -- usually to the US -- reinforces instability, uncertainty and risk-premium. Financial markets are, therefore, an important channel of the US monetary policy. 

The magnitude and the overall dynamics of the international reaction depend on how the US economy reacts to monetary policy shocks. A crucial aspect, as covered in the seminal contribution by \citet{bloom2009impact}, is the nature of the shock.  A first-moment shock typically elicits a more sustained macroeconomic response (e.g., in output), while an uncertainty shock results in a quicker, more pronounced drop, followed by a rebound.  That means that impact on emerging markets also differs depending on whether we allow for the first-moment, second-moment, or both types of shocks in the US economy. In fact, as elaborate in \citet{Leduc2016} and \citet{basu2017uncertainty}, a volatility shock reminds a demand shock but the shape and dynamics of outcome variables are different, requiring a joint framework to account for the standard monetary policy shock and changes in uncertainty.

We contribute to the literature threefold. First, we produce a global empirical macroeconometric model covering almost the entire world -- nearly $90\%$ of the world output -- to quantify the direct and indirect impacts of the US monetary policy change on the emerging markets. As alluded to earlier, this aspect is particularly important for less directly connected countries since the US impact may stem from other affected economies. Our model allows country-specific reactions to domestic as well as global variables. Second, much of the literature focuses on the monetary policy shocks impact or shocks to the monetary policy uncertainty; we complement with an analysis on distinguishing between indirect and direct spillover effects by considering the role of monetary policy uncertainty on global impacts when the Fed unexpectedly tightens monetary policy, i.e., the first-moment tightening shock is associated with an increase in uncertainty on the future path of monetary policy. This situation is, in fact, reminiscent of the high inflation environment of 2022-2023, when the central bank deals with banking sector volatility, financial stability tradeoffs, and the still unclear path of the macroeconomy. 
To our best knowledge, it is the first paper to quantify spillover strengths, both  direct and indirect, while taking into account changes in (policy and macro) uncertainty. Last but not least, in line with the empirical DSGE model approximation as implemented by \citet{Dees2014}, we also consider an explicit role for the stock market, with a focus on emerging markets. We can therefore shed fresh light on the existence of the global financial cycle driven by the US monetary policy.  

We show that, as expected, the Fed's tightening of monetary policy reduces output growth and inflation, but what is less covered in the literature is that it leads to an increase in US macroeconomic volatility, especially inflation volatility, therefore supporting the findings of \citet{mumtaz2019dynamic}. Furthermore, we show that when a  positive US interest rate shock is complemented by a simultaneous Fed's monetary policy uncertainty, output and prices respond more strongly, confirming that the first-moment shock experienced with an increase in the second moment leads to deeper recessions (see \citealp{BloomForthcoming}). That suggests that agents respond to the unpredictability of the central bank's reaction function, creating a link between the macroeconomy's fundamentals and policy uncertainty. Therefore, our work is related to the existing literature that uses external sources of uncertainty and demonstrates that monetary policy, which may require larger space to accommodate a deeper decline, becomes less effective in a (predetermined) state of high uncertainty (e.g., \citealp{CP2018,Pellegrino2018}), thus lending support to the real-options effects, with fresh insights when policy may actually be strengthened. What we document is that the model-consistent monetary policy uncertainty looks like a negative demand shock (see, e.g., \citet{Leduc2016,BloomForthcoming} since it reduces GDP (increases unemployment), makes prices (inflation) go down, and increases agents' cautiousness. Hence, an increase in interest rates, once associated with a rise in monetary policy uncertainty, can reinforce each other, making output and inflation drop more strongly. Furthermore, volatilities of output and inflation rise substantially more, causing agents to postpone decisions or behave more cautiously. 

Importantly, our paper shows strong and heterogeneous spillover effects of US monetary policy rate hike that lead to declines in emerging markets' output growth, inflation, and exchange rate depreciation, especially when policy change is associated with a simultaneous surge in policy uncertainty. The real variables (output) react more strongly for nearby emerging economies (Latin America in comparison to Asia), indicating the role played by the demand channel. 
In addition, we find that asset price adjustments are significantly larger when uncertainty increases, hinting about the importance of financial uncertainty in generating business cycles, as recently found by \citet{ludvigsonuncertainty}. What is more, we provide additional empirical support for the global financial cycle phenomenon (\citealp{Rey2015, Miranda-Agrippino2020, MirandaAgrippino2020}) by documenting stock market responses, particularly prominent in Latin America. 
Finally, we show the evidence of spillovers occurring not only via direct (bilateral) linkages with the US but also via indirect ones. Both channels are amplified when first and second-moment shocks occur together. China's inflation and renminbi are particularly sensitive to US monetary policy when indirect effects are taken into account, pointing to the role of Chinese integration into the global economy. 

Our paper is structured as follows: Section \ref{sec:Literature} reviews and places our paper within the related literature. The empirical model and shock identification schemes are explained in Section \ref{sec:empirical}. Next, empirical findings of the baseline model are first described in Section \ref{sec:Results}, while Section \ref{sec:Discussions} offers more detailed discussions and extensions with the equity prices included in the empirical model and more details are provided about the decomposition of direct and indirect effects. Finally, Section \ref{sec:Conclusions} provides policy implications and a few concluding remarks.

\section{Related Literature}\label{sec:Literature}

Our study mainly relates to two main streams of the literature. First, we provide insights into US monetary policy spillovers into emerging markets. The literature on the global effects of the US monetary policy shocks is vast (see, among others, \citealp{breitenlechner2021goes,CaZorzi2021,crespo2019spillovers,Dees2021,Degasperi2020,Georgiadis2016,MirandaAgrippino2020,Pinchetti2021}). The classical channels of contractionary monetary policy include the demand (trade) effect, which makes global output co-move, and, relatedly, the exchange rate depreciate (dollar appreciation), thereby leading global demand move away from the US to foreign goods, and the financial channel (changes in interest rates and equity prices). The demand channel operating through the exchange rate adjustments (see, e.g., \citealp{breitenlechner2021goes,CaZorzi2021,Cesa-Bianchi2022,Dees2021}) is especially important for emerging markets. However, the financial channel is also becoming more prominent due to emerging markets deeper financial integration (\citealp{Ahmed2021,Dees2021,Degasperi2020,Feldkircher2016,crespo2019spillovers}). In our model we will account for demand and trade linkages (also allow for the exchange rate to respond to changes in the US monetary policy) as well as explore changes in interest rates and asset prices (the latter is added in the extended model). Effectively, we will confirm the importance of the US dollar status and the existence of the global financial cycle (\citealp{Giovanni2021,Miranda-Agrippino2020,MirandaAgrippino2020,Rey2015,LN:2021}). 

Interestingly, since some emerging markets, in particular, China, has accumulated huge reserves of the US dollar, it also became more sensitive to changes in the US economy. Among key financial channels are the investment channel (\citealp{Avdjiev2019}) and the credit channel (\citealp{McCauley2015}), emphasizing the importance of financial conditions, in particular, the US interest rates. The relevance of exchange-traded funds as a conduit of cross-border capital flows and thus an increased exposure of emerging markets to the global financial cycle is documented by \citet{Converse:2020}. In the extension of our model, we analyze how domestic equity markets react to changes in the US monetary policy stance. 

Secondly, we contribute to the literature on uncertainty and its impact on the magnitude of spillovers. The focus on emerging markets is often motivated by their fragility and the non-negligible spillback effects to more advanced economies. For example, \citet{Fernandez-Villaverde2011} found that interest rate volatility shocks have a significant effect on output and act as a critical driver of business cycle fluctuations in Latin America. An increased global financial uncertainty among advanced and emerging small open economies is a crucial driver of output downturns, especially for countries with higher levels of vulnerability. Our study sheds new light on how the advanced economy, facing greater uncertainty, shapes economic dynamics in emerging markets. The importance of these markets has grown due to their increased weight in the global economy and the risk of adverse spillbacks (\citealp{BIS2016,Agenor2018,Carney2019}). And our emphasis on the uncertainty about the future path of US monetary policy is different from the existing literature that focuses on emerging market domestic vulnerabilities (e.g., \citealp{Ahmed2021,Georgiadis2016,Ruch2020}). 

A recent exception is \citet{Arbatli-Saxegaard2022}, which demonstrates that the source of the change in US interest rates also matters in addition to local conditions. The effects of US monetary policy shocks are not limited to economic activity but also extend to the distribution of macroeconomic outcomes, with a stronger impact on the lower end of the distribution. Another paper looking at the ECB's monetary policy effectiveness when the euro area faces more uncertainty is by \citet{Pellegrino2018}. It is shown that monetary policy is less effective when Europe faces more uncertainty. Our analysis looks at identified monetary policy, rather than general, uncertainty. It also tracks distributional changes in the sense that higher moments of macroeconomic variables are affected. In other words, we show that US monetary policy tightening can be significantly amplified once accompanied by a simultaneous increase in uncertainty on its future path, therefore generating higher levels of uncertainty globally thus impacting the size and shape of spillovers. We also emphasize indirect channels when headwinds are not only blowing through direct exposure to the US economy but also through other countries impacted by the Fed's policy changes.

The spillover analysis joining the first and second moment of monetary policy shocks is inspired by \citet{BloomForthcoming}, and we confirm that a shock to the policy rate complemented with the second-moment shock is more recessionary and disinflationary. Moreover, interacting with the trade linkages can deliver statistically and economically significant impacts on emerging markets. Our analysis is, therefore, also related to the recent literature that explores shocks to uncertainty. For us, the key focus is on the US monetary policy shock transmission and spillovers to the emerging markets, conditional on different levels of policy uncertainty. In the vast literature on domestic uncertainty (\citealp{BloomForthcoming,Caggiano2014,creal2017monetary,Fasanietal,Fernandez-Villaverde2020,Giavazzi2012,husted2019monetary,Leduc2016,ludvigsonuncertainty,mumtaz2018policy}) and global (\citealp{BHATTARAI2019,Biljanovska2021,Bonciani2020,Cesa-Bianchi:2019aa,Chinn:2017aa,Colombo2013,cuaresma2020fragility,Lakdawala2021,LN:2021}), an emphasis is on how the second-moment shock differs to the first-moment shock (in terms of the channels, impacts, transmission mechanisms, etc.). We provide evidence of how a shock to US monetary policy uncertainty changes the transmission of the first-moment shock, i.e., monetary policy tightening. 

\section{Empirical Model} \label{sec:empirical}

\subsection{Introducing the GVAR-SV Model}

We base our modeling strategy on the new empirical model, recently contributed by \citet{LN:2021}. It is a global vector-autoregression with endogenous stochastic volatility (GVAR-SV) model that can track how monetary policy spills from the US to the global economy. We extend the GVAR-SV model's use by considering the conventional, first-moment shock to the US monetary policy and its impact on emerging economies, distinguishing between direct and indirect effects as well as considering the environment that admits and disregards uncertainty. In contrast, the focus of \citet{LN:2021} was on the transmission of the second-moment US monetary policy shock. As detailed in \citet{Dees2014}, and what constitutes our baseline specification, the approximation of the non-linear DSGE model entails a four-variable system: the Euler equation for output, the short-term interest rate, the real exchange rate, and inflation, all augmented with the weighted averages of main variables to capture global interdependencies. Following \citet{LN:2021}, we introduce the stochastic volatility terms for each endogenous variable. Our extension also includes the role of financial markets; thus, we augment the model with the fifth variable, equity prices. It will help us track how real and financial variables in the emerging markets respond to the unexpected monetary policy tightening in the US.

Notably, a crucial element of the model is the  stochastic volatility block, which helps us identify how uncertainty reacts and impacts macroeconomic developments as well as strengthens or impedes monetary policy transmission. Technically, we build on the seminal contributions of \citet{M.H.Pesaran2004,Dees2014} who proposed and applied the global vector autoregressive model (GVAR), which enables tracking the global economy by using cross-sectionally weighted averages (so-called foreign variables) and thus reducing the dimensionality of the estimation problem.\footnote{The GVAR methodology has seen substantial growth, both theoretically and in terms of applications. For instance, \citet{ChudikPesaran2011,ChudikPesaran2013} establish the conditions under which the key macroeconomic variables can be arbitrarily well approximated by a set of finite-dimensional small-scale models that can be consistently estimated separately and then stacked into the global model. An alternative route would be to apply Bayesian techniques, as covered by \citet{banbura2010large,huber2016density}. However, the GVAR methodology affords economic interpretability by using real-world trade (or other) linkages.} Writing out a model for each economy, expressed as the vector autoregression with weakly exogenous variables (VARX) and augmenting it with stochastic volatility, we obtain:

\begin{equation}
\boldsymbol{x}_{it}=\boldsymbol{a}_{i}+\sum_{\ell=1}^{p_{i}}\boldsymbol{\Phi}_{i\ell}\boldsymbol{x}_{i,t-\ell}+\sum_{\ell=0}^{q_{i}}\boldsymbol{\Lambda}_{i\ell}\boldsymbol{x}_{i,t-\ell}^{*}+\sum_{\ell=0}^{s_{i}}\boldsymbol{\Psi}_{i\ell}\boldsymbol{h}_{i,t-\ell}+\boldsymbol{u}_{it},\label{VARX}
\end{equation}
where $\boldsymbol{a}_{i}$ denotes vector of intercepts, $\boldsymbol{x}_{it}=[x_{i1t},..,x_{ik_{i}t}]'$ denotes
the $k_{i}\times1$ vector of domestic variables, 
$\boldsymbol{\Phi}_{i}$ is the coefficient matrix associated with the lags of $\boldsymbol{x}_{it}$, $\boldsymbol{x}_{it}^{*}$
denotes the $k_{i}^{*}\times1$ vector of foreign variables in country
$i$ associated with the coefficient matrix $\boldsymbol{\Lambda}_{i}$. 

As already alluded to above, and following the literature on empirical open-economy DSGE models (\citet{Clarida2001,Gali2005,Dees2014}), $\boldsymbol{x}_{it}$ includes annual output growth, inflation, short-term interest rate, and the real exchange rate growth, whereas $\boldsymbol{x}_{it}^{*}$ comprises respective cross-sectional averages to capture common unobserved factors (or multiple factors) (see \citealp{Pesaran2006}). The extension undertaken by \citet{LN:2021} also augments the traditional VARX models with endogenous stochastic volatility equations: $\boldsymbol{h}_{it}=\left[h_{i1t},h_{i2t},...,\,h_{ik_{i}t}\right]$,
where $\boldsymbol{\epsilon}_{it}$ is a $k_{i}\times1$ vector of log volatility of structural shocks. The VARX residuals $\boldsymbol{u}_{it}$ are assumed to have  a time-varying variance-covariance matrix $\boldsymbol{\Omega}_{it}$: $V(\boldsymbol{u}_{it})=\boldsymbol{\Omega}_{it}$. The relationship between the residuals $\boldsymbol{u}_{it}$ and the structural shocks $\boldsymbol{\epsilon}_{it}$ is assumed to be as follows:
 \begin{equation} \label{eq:residual-structural}
\boldsymbol{A_{i}} \boldsymbol{u}_{it} = \boldsymbol{\epsilon}_{it},
\end{equation}
where the form of matrix $\boldsymbol{A_{i}}$ captures the contemporaneous relationship among the reduced-form shocks and helps identify structural shocks $\boldsymbol{\epsilon}_{it}$. Next, $\boldsymbol{\epsilon}_{it}$ represents structural shocks with a covariance matrix $\boldsymbol{H}_{it}$, i.e., $V(\boldsymbol{\epsilon}_{it})=\boldsymbol{H}_{it}$, which is a diagonal matrix whose values are the volatility of structural shocks.
Eq. (\ref{eq:residual-structural}) can be also written as: 
 \begin{equation} \label{eq:residual-structural-v2}
\boldsymbol{A_{i}} \boldsymbol{u}_{it} = \boldsymbol{H}_{it}^{1/2} \boldsymbol{e}_{it}.
\end{equation}
where $\boldsymbol{e}_{it}$ is the standardized structural shocks, $\boldsymbol{e}_{it}\sim N(\boldsymbol{0},\,\boldsymbol{I})$, and  $\boldsymbol{\epsilon}_{it}=\boldsymbol{H}_{it}^{1/2} \boldsymbol{e}_{it}$.  
Based on Eq. (\ref{eq:residual-structural-v2}), the variance-covariance matrix $\boldsymbol{\Omega}_{it}$ can be expressed as:
 \begin{equation} \label{eq:residual-structural-variance}
\boldsymbol{\Omega}_{it}= \boldsymbol{A_{i}}^{-1} \boldsymbol{H}_{it} \boldsymbol{A_{i}}^{-1\prime}.
\end{equation}

Unlike a standard assumption in the literature that the stochastic volatility is  subject to an autoregressive process (i.e., it is an exogenous force impacting macroeconomic variables), we express the transition equation as:
\begin{equation}
\begin{split}\boldsymbol{h}_{it}=\boldsymbol{c}_{i}+\sum_{\ell=1}^{m_{i}}\boldsymbol{\Upsilon}_{i\ell}\boldsymbol{h}_{i,t-\ell}+\sum_{\ell=1}^{q_{i}}{\boldsymbol{\Xi}}_{i\ell}\boldsymbol{x}_{i,t-\ell}+\boldsymbol{\eta}_{it},\\
\quad\ensuremath{\boldsymbol{\eta}_{it}\sim N(\boldsymbol{0},\,\boldsymbol{Q}_{i})}\,\textrm{and}\,\ensuremath{\mathbb{E}(\boldsymbol{e}_{it},\,\boldsymbol{\eta}_{it})=\boldsymbol{0}},
\end{split}
\label{eq:h_dynamics}
\end{equation}
where $\boldsymbol{c}_{i}=\left[c_{i1},...,\,c_{ik_{i}}\right]^{\prime}$
is a vector of intercepts, $\boldsymbol{\Upsilon}_{i}$ is a matrix
of coefficients on the lagged log volatility of structural shocks,
and ${\boldsymbol{\Xi}}_{i}$ captures the effects of lagged
macroeconomic variables on log volatility of structural shocks in
country $i$. The volatility shocks  $\boldsymbol{\eta}_{it}$ and the structural level shocks $\boldsymbol{e}_{it}$ are uncorrelated and $\boldsymbol{Q}_{i}$, i.e. the covariance matrix of  $\boldsymbol{\eta}_{it}$, is assumed to be diagonal. And the final ingredient to construct the global model (GVAR-SV) is weighting of the endogenous variables to produce the so-called foreign variables $\boldsymbol{x}_{it}^{\ast}$ in (\ref{VARX}). They approximate the unobserved factors:
\begin{equation}
\boldsymbol{x}_{it}^{\ast}=\bar{\boldsymbol{W}}_{i}\boldsymbol{x}_{t},\label{eq:weighting}
\end{equation}
where elements of matrix $\bar{\boldsymbol{W}}_{i}$ are given by
$w_{ii}=0$ and $\sum_{j=0}^{N}w_{ij}=1$ for all $i,\,j=0,\,1,\ldots,\,N$ and $\boldsymbol{x}_{t}=[\boldsymbol{x}_{1t}^{'},..,\boldsymbol{x}_{Nt}^{'}]'$ is the set of $k=\sum_{i=0}^{N}k_{i}$ already introduced endogenous variables. 

These connections allow us to investigate the spillover effects of shocks. To see how the US monetary policy shock's spillover effects are quantified, consider the following example. An unexpected shock in country $i$ first affects $\boldsymbol{x}_{it}$ via (\ref{VARX}), then spillovers to countries $j\neq i$ via their foreign specific variables $\boldsymbol{x}_{jt}^{\ast}$, and spillbacks to country $i$ via $\boldsymbol{x}_{it}^{\ast}$. Therefore, the model captures both the (bilateral) direct impacts from the shock country of origin $i$ to country $j$ as well as the indirect impacts via linkages of country $j$ (different from $i$). The total spillovers will be the sum of both direct and indirect impacts. There is a possibility that the direct effect is small (given no direct linkages between the two countries) while the indirect impact is large because other connected partners are hit hard. This aspect is vital for emerging markets that may not be impacted directly by the actions in the US but nevertheless experience the strain. It is important to note that each foreign variable is country-specific, reflecting exposure to the global economy through trade flows, and each parameter is also unique to every covered economy. These aspects are important for the set of diverse emerging markets since pooling can lead to poor and even inconsistent inference (see \citealp{Pesaran1995}). 

It is worthwhile to draw attention to the fact that emerging markets are prone to a more volatile economic environment, susceptible to changes in the global economy, particularly the US.  And since the contemporaneous and lagged log volatility of structural shocks $\boldsymbol{h}_{it}$ are featured in (\ref{VARX}), we can explore how macroeconomy adjusts to policy shocks as well as changes in uncertainty.\footnote{Our empirical model resembles a reduced form of a DSGE model approximated to the third order around the steady state, even if quadratic and cubic terms are missing. As argued by \citet{mumtaz2015international}, the second moments are sufficient to capture the  responses of macroeconomic aggregates to uncertainty.} Since unobserved factors with country-specific factor loadings are featured in the macroeconomic variables, which, in turn, impact volatility, we allow for a very rich dynamics between fundamental and uncertainty in the global setting.\footnote{The full solution of the global GVAR-SV model is presented in \citet{LN:2021}.}

\subsection{Identification}

 The VARX for the US comprises three standard endogenous variables
(interest rate, output growth and inflation) and three exogenous
variables, namely, foreign output growth, foreign inflation and
foreign exchange rate changes. We identify the US monetary policy shocks by sign restrictions, following \citet{mumtaz2013impact}. Specifically, we consider the following structure for $\boldsymbol{\tilde{A}}=\boldsymbol{A}^{-1}$ (for brevity we drop the country notation $i$ in $\boldsymbol{A}_{i}$):
\[
\boldsymbol{\tilde{A}}=\begin{pmatrix}1 & 0 & 0\\
a_{2,1}^{(-)} & 1 & 0\\
a_{3,1}^{(-)} & a_{3,2} & 1
\end{pmatrix},
\]
where the superscript $(-)$ denotes the negative sign restrictions
in the corresponding parameters. 

The ordering of the endogenous variables are interest rate, output growth, and inflation, which implies that a shock to interest rate has a contemporaneous impact on GDP growth and inflation as supported by New Keynesian models. For monetary policy shock identification, we also apply the sign restrictions. Particularly, an increase in interest rates causes a contemporaneous fall in output growth and inflation, an assumption which is implied by standard DSGE models and commonly adopted in the literature. The sign restrictions are helpful to avoid issues such as the price puzzle. If we only impose sign restrictions without the recursiveness assumption on the structure of $\boldsymbol{\tilde{A}}$, the draw of elements of $\boldsymbol{A}$ would become rather cumbersome. Note that unlike a standard VAR model, $\boldsymbol{\tilde{A}}$ cannot be rotated after estimation to impose sign restrictions. This is because the log volatility enters the VAR equations; hence, changes to $\boldsymbol{\tilde{A}}$ have an impact on the stochastic volatility which, in turn, affects the VAR coefficients. 

\begin{figure}[!h]
\raggedright{}\caption{\label{fig:Impacts-of-USNO-level}Impacts of US Monetary Policy Shocks}
\hspace*{-0.5in} \includegraphics[width=0.8\paperwidth]{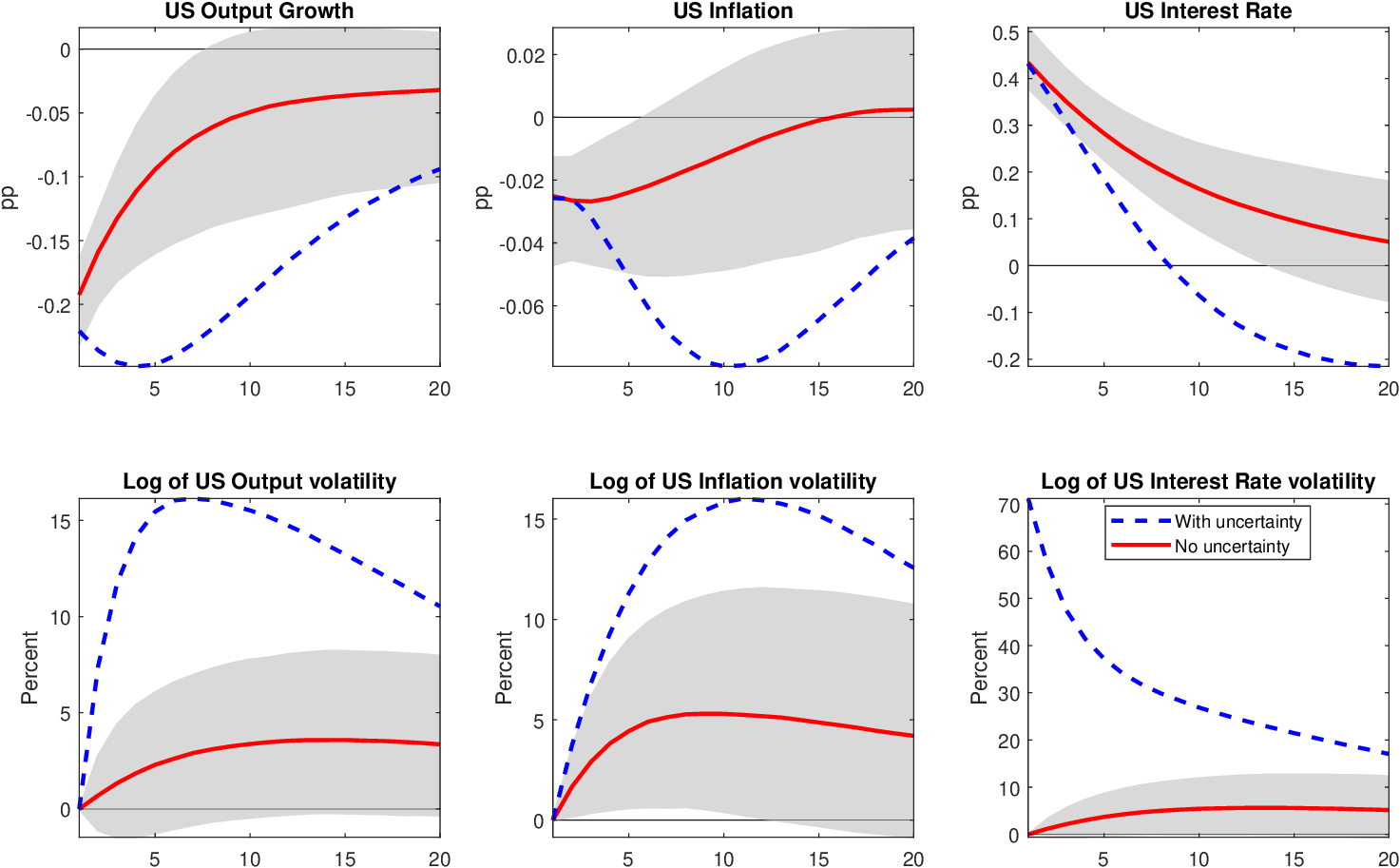}
 \begin{flushleft}
   \footnotesize{Notes: Figure presents the response of US macroeconomy to one standard deviation shock to the US interest rate: in each entry, the red solid line and the shaded area are the median response and the 68 percent intervals where there is \textit{no} shock to the US monetary policy uncertainty (MPU); the blue dashed line is the response when one standard deviation shock to the US interest rate occurs with one standard deviation shock to the US policy uncertainty.}
 \end{flushleft}
\end{figure}

In our analysis, we concentrate on the spillovers of an unexpected increase in the US interest rate, separating the direct versus indirect linkages. To highlight the role played by uncertainty, we investigate two cases: the first one is with only US monetary policy shock, keeping volatility shocks shut, and the second one also allows for a simultaneous increase in the associated volatility of US monetary policy shocks in the spirit of \citet{bloom2009impact}. 
While the increase of US interest rate in both cases is the same (equal to the average one standard deviation over the sample, see more the discussions below and in Figure \ref{fig:Impacts-of-USNO-level}), we will show that allowing for uncertainty within the same global model also matters.

\subsection{Data}

The global model developed in this paper includes the US along with other 32 countries, accounting for $90\%$ of world output including both advanced Economics and emerging markets: Australia, Canada, Japan, New Zealand, Norway, Singapore, Sweden,  Switzerland, UK, and 8 euro area (EA) countries, namely Austria,  Belgium, Finland, France, Germany, Italy, Netherlands, and Spain. The latter includes emerging Asian countries: China, India, Indonesia, Malaysia, Philippines, South Korea, and Thailand; emerging Latin American economies: Argentina, Brazil, Chile, Mexico, and Peru and  other emerging economies: Turkey, Saudi Arabia, and South Africa. In line with the GVAR literature, eight countries that originally joined the euro on 1 January 1999 are grouped together using the average Purchasing Power Parity GDP weights. The GVAR model hence contains 26 countries/regions modeled individually.

For non-US VARX model, we include four endogenous variables for each country/region: annual output growth, inflation, short-term interest rate and the real exchange rate growth. The exchange rate does not enter the US model since the US dollar is used as a num\'eraire.\footnote{The data are available from the GVAR quarterly database; see \citet{mohaddes2018compilation} for details.} For the US, the euro area, the UK, and Japan, we use the shadow rate from Morgan Stanley, which remains a useful measure of monetary policy stance at the effective lower bound.

To construct the country-foreign specific variables, in the baseline model we use the fixed trade weights, which are the average trade flows computed over the years 1990-2016. In line with the GVAR literature, for instance \citet{dees2007exploring}, excepting the US, all models include three country-specific foreign variables: foreign output growth, foreign inflation and foreign interest rate, as weakly exogenous. In the case of the US model, we include foreign output growth, foreign inflation and weighted average of bilateral real exchange rates, i.e., the real effective exchange rate, as weakly exogenous. Given the importance of the US financial variables in the global economy, the US-specific foreign interest rate is not included in the US model. In contrast, the US-specific foreign output growth and inflation variables are included in the US model in order to capture the possible second-round effects of external shocks on the US.

The estimation of the models spans from 1979Q2 to 2019Q4, utilizing the period between 1979Q2 and 1989Q2 as a training sample for constructing priors, in line with the methodologies of \citet{cogley2005drifts} and \citet{mumtaz2015international}. For each country, we implement a Bayesian approach to estimate the VARX system, a non-linear state space model comprising equations (\ref{VARX})-(\ref{eq:h_dynamics}) and the definition of cross-sectional averages given in equation (\ref{eq:weighting}).\footnote{See \citet{LN:2021} for details on the initial condition and estimation algorithm of the model.}

\section{\label{sec:Results}Main Findings}

We start by subjecting the US economy to the identified monetary policy shock, seeking to quantify the impact of an unexpected rise in the interest rate. Figure \ref{fig:Impacts-of-USNO-level} visualizes macroeconomic responses in output growth, inflation, interest rate, and the associated volatilities (the upper panel refers to the macroeconomic variables, whereas the lower panel depicts the volatilities of the same variables.). We depict two impulse response functions -- which, in the conditional expectation function sense, show the predicted variable of interest under a monetary policy shock and without it. Focusing on the red lines, which represent median responses to the Fed monetary policy tightening, we see that our model replicates what a standard macroeconomic model would suggest: upon an unexpected increase in the US interest rate, output growth declines, and so do prices. 

Quantitatively, one standard deviation shock to the US interest rate (around 0.4 pp) leads to 0.2 pp drop in output and 0.02 pp drop in inflation on impact. The output returns to the pre-shock level very slowly, whereas inflation is somewhat faster to adjust. In line with the literature on uncertainty (e.g., refer to \citet{basu2012uncertainty,bloom2009impact,BloomForthcoming, mumtaz2019dynamic} for theoretical arguments and empirical evidence), monetary policy tightening coupled with a rise in interest rate volatility leads to substantially deeper drops in output and inflation, generating important policy implications.

\subsection{US Monetary Policy Tightening and Monetary Policy Uncertainty}

Interestingly, even when exploring the first-moment shock, we can generate ensuing inflation uncertainty (refer to the lower panel). In other words, an unexpected rise in monetary policy tightening is associated with higher interest rate volatility. And even though output and interest rate volatilities are not statistically significantly elevated when monetary policy uncertainty is not shocked, uncertainty about inflation, as measured by its volatility, increases significantly (refer to the solid line). Therefore, we document an empirical fact that it becomes harder to make decisions that rely on the prediction of inflation. Clearly, a more uncertain macroeconomic environment feeds back into the Fed's decision-making process, as it directly impacts the Fed's monetary policy effectiveness. 

Given that our model captures reasonable macroeconomic dynamics, we now make use of its richness by subjecting the US macroeconomy to two shocks at the same time -- Fed's monetary policy tightening and monetary policy uncertainty.  In other words, not only does the interest rate unexpectedly increase, but so does a monetary policy shock variance. Unsurprisingly, then, once we take a shock to the US interest rate uncertainty into account (dashed blue line), although interest rate increases by the same amount on impact, output drops by almost 0.3 pp after 5 quarters, whereas inflation is lowered by approximately 0.08 pp after 10 quarters. With a more recessionary and more disinflationary impact, interest rate returns back faster and may later drop after an initial rise, reflecting a Taylor-rule-like behavior. That implies that agents react to the fact that the central bank's reaction function is less predictable, inducing an endogenous relationship between the macroeconomy (fundamentals) and policy uncertainty. In other words, once we allow for shocks to the interest rate and its volatility, not only the macroeconomic impacts get amplified, but also US output and inflation volatilities increase substantially.

Unlike a standard exercise with monetary policy tightening only, a change in both interest rate level and volatility makes the whole macroeconomic environment less predictable, thereby impacting policymakers and all other economic actors. And indeed, following the logic of \citet{bloom2009impact,BloomForthcoming}, business cycle fluctuations can be rationalized by high uncertainty -- economic activity slows down due to wait-and-see and precautionary motives (\citealp{Bachmann2013,Bloom2007,Stokey2016}). A pause in investment and new employment brings productivity down, thus generating a sharp recession. Importantly for our paper, \citet{jurado2015measuring} emphasize that uncertainty has large and persistent impacts on the economy, with significant rises in recessions (\citealp{bloom2009impact, jurado2015measuring}), partly due to monetary policy shocks  (\citealp{mumtaz2019dynamic}). And recently \citet{ludvigsonuncertainty} show that the source of uncertainty matters for the endogenous effects, emphasizing the interplay between changes in the economic fundamentals and uncertainty. We set out to explore not only how the macroeconomy endogenously responds to monetary policy changes when uncertainty rises but also how it spills to emerging markets. The GVAR-SV model is well suited for that purpose as it can shed light on the strength of spillovers under different monetary policy environments in the US.

\subsection{US Monetary Policy Tightening and Spillovers}

Given the results of our model, we can now answer the question quoted in the introduction: indeed, when the US sneezes, or its monetary policy gets tightened, a cold is caught or the Fed's policy effects get spilled to emerging markets. There is a direct effect due to the dollar's global status and direct linkages, and other effects reaching from affected foreign markets. We concentrate our attention to the emerging markets group (Argentina, Chile, India, Korea, Malaysia, Philippines, South Africa, and Thailand), then China, then emerging markets Asia without China, and emerging markets Latin America (Argentina and Chile). These groups allow tracking important heterogeneities how the US monetary policy gets transmitted to some of the most systematically important emerging countries.

\begin{figure}[H]
\raggedright{}\caption{\label{fig:Impacts-of-USNO-level-ALL} Impacts of US Monetary Policy Shocks: Spillovers}
\hspace*{-0.5in} \includegraphics[width=0.8\paperwidth,height=0.9\paperwidth]{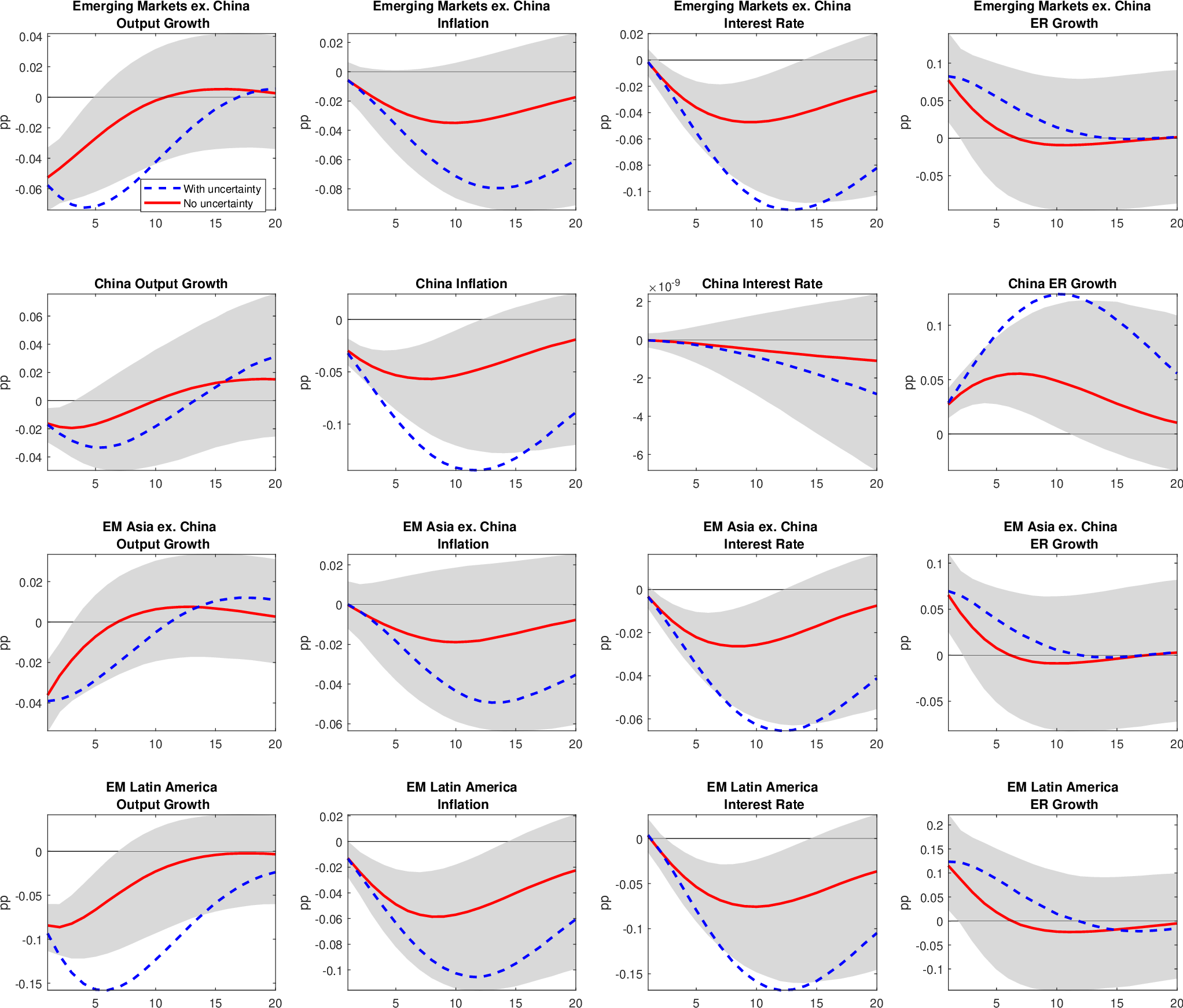}
 \begin{flushleft}
   \footnotesize{Notes: Figure presents the response of emerging markets to one standard deviation shock to the US interest rate: in each entry, the red solid line and the shaded area are the median response and the 68 percent intervals where there is \textit{no} shock to US MPU; while the blue dashed line is the response when one standard deviation shock to the US interest rate occurs with a one standard deviation shock to the US policy uncertainty.}
 \end{flushleft}
\end{figure}

First, we document a global slowdown effect because of the Fed's monetary policy tightening (Figure \ref{fig:Impacts-of-USNO-level-ALL}). The output growth declines across the board, with the largest negative contemporaneous impact in Latin America and the most modest one in China (0.08 and 0.02 pp, respectively). The demand -- or trade channel -- hints that gravity forces work, and the geographically closest group's output adjusts most. The picture is slightly different when we analyze financial or nominal variables. The monetary policy tightening is disinflationary. However, the effect is statistically significant for the Chinese and Latin American economies, whereas impact on prices in other Asian emerging markets lacks precision. Even though inflation is not impacted significantly everywhere, since output growth declines, interest rates get lowered across the board. That has important policy implications -- a global slowdown induced by the Fed's monetary policy tightening makes emerging economies' central banks grapple with the consequences. We will discuss policy implications in more detail in the Section \ref{sec:Conclusions}.

The emerging markets' central banks attempt to revive their economies by engaging in accommodative monetary policy. Unsurprisingly, then, that local currencies versus the US dollar depreciate on impact across the board. Though the effect is short lived in all country groups, that is not true for the Chinese economy. While interest rate appears not responding in order to stabilize the economy, the renminbi stays at higher levels for sufficiently longer time, which helps improve country's export competitiveness and boost output growth, a finding that is in line with  arguments made by several authors (e.g., \citealp{Steinberg2015, Mattoo2017}).

Turning attention to the spillovers when we allow US monetary policymakers to face elevated uncertainty, we find similar but stronger effects. Our results empirically confirm \citet{BloomForthcoming}, documenting that recessions are best modelled as being driven by first-moment (in our case, interest rates) shocks and a positive second moment (in our case, interest rate volatility). For the overall emerging markets group without China, as well as for Latin America, output declines more and that necessitates a larger drop in interest rates, when US monetary policy uncertainty hikes. Reactions in inflation and exchange rates do not significantly differ when we consider a standard case without changes in uncertainty. 

The story is, however, different for China, implying that the US impact on China operates differently than on other emerging markets. Chinese prices fall substantially more, and the depreciation of the renminbi is larger, with the latter making its exports cheaper and thus mitigating the negative impact on output. Our results also suggest a weak stabilization channel via the interest rate instrument in China. Lastly, when we look at the Asian economies excluding China, an increase in the US interest rate and its uncertainty leads to a larger drop in domestic interest rate than compared to the case when only the Fed's monetary policy tightening occurred. Before extending the model, we conclude that the demand and price channels seem more important for Latin American and Chinese economies, respectively.

\begin{figure}[!h]
\raggedright{}\caption{\label{fig:Impacts-of-US-Equity}Impacts of US Monetary Policy Shocks: Equity Price}
\hspace*{-0.5in} \includegraphics[width=0.8\paperwidth, height=0.5\paperwidth]{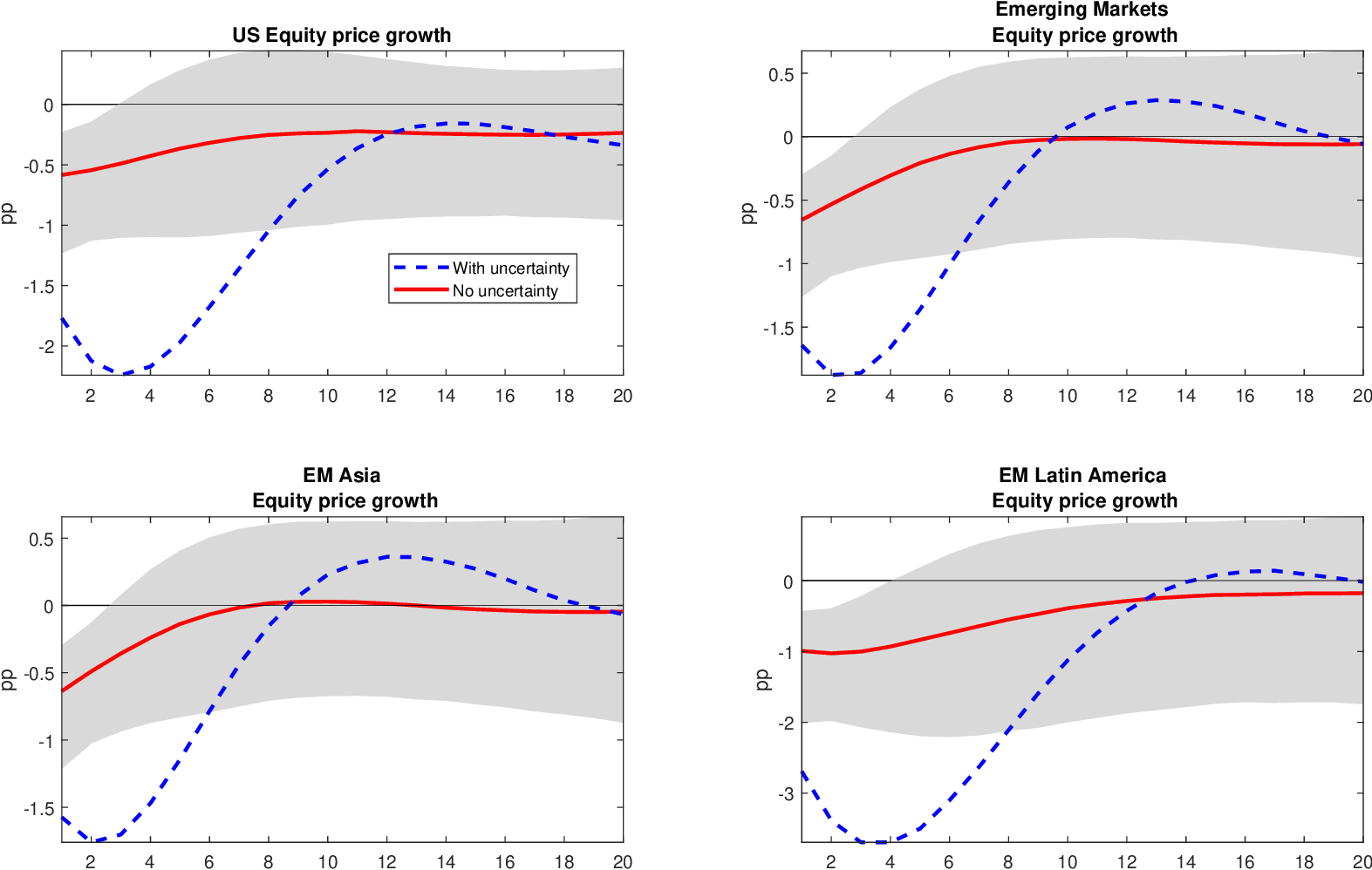}
 \begin{flushleft}
   \footnotesize{Notes: Figure presents the response of equity price to one standard deviation shock to the US interest rate: in each entry, the red solid line and the shaded area are the median response and the 68 percent intervals where there is \textit{no} shock to US MPU; while the blue dashed line is the response when an 1 standard deviation shock to the US interest rate occurs with one standard deviation shock to the US policy uncertainty. Emerging markets response is weighted average response of countries where data is available: Argentina, Chile, India, Korea, Malaysia, Philippines, South Africa, and Thailand. EM Asia response is weighted average response of India, Korea, Malaysia, Philippines, and Thailand. EM Latin America response is weighted average response of Argentina and Chile.}
 \end{flushleft}
\end{figure}

\section{Discussion}\label{sec:Discussions}

\subsection{Impacts of US Monetary Policy Shock on Equity Price}\label{sec:equity}

Our baseline model is expanded to include equity prices. Figure \ref{fig:Impacts-of-US-Equity} provides evidence for the existence of global financial cycle. This extension is important since adjustment in capital flows due to changes in investors' risk perceptions is one of key channels how emerging markets get affected by tightening US monetary policy.  That leads to flight to safety (or flight to quality),   as recently evidenced by \citet{Avdjiev2019}, documenting lower cross-border bank flows and lower investment in emerging markets and changes in dollar credit (\citealp{Giovanni2021,McCauley2015}). A contractionary interest rate shock makes US asset prices decline. However, the reaction is relatively modest, whereas when we consider a situation when uncertainty unexpectedly increases, a contemporaneous effect increases almost fourfold. Similarly, Asia's, Latin America's and the overall emerging countries group equity prices get depressed when the US tightens monetary policy. 

The effect on equity prices is considerably larger when the US faces uncertainty, exemplifying the importance of the intricate link between uncertainty, monetary policy, and the real economy (\citealp{baker2016measuring,jurado2015measuring,ludvigsonuncertainty}). In other words, in the context of increasing US interest rates, drops in equity prices and the ensuing volatility of financial markets necessitate policy actions by central banks, financial regulators, and sometimes international bodies such as IMF, especially for emerging economies with substantial external financing requirements.

\begin{figure}[!h]
\raggedright{}\caption{\label{fig:Impacts-of-USNO-level-US-Direct}Impacts of US Monetary Policy Shocks: Direct versus Total Spillovers}
\hspace*{-0.5in} \includegraphics[width=0.8\paperwidth]{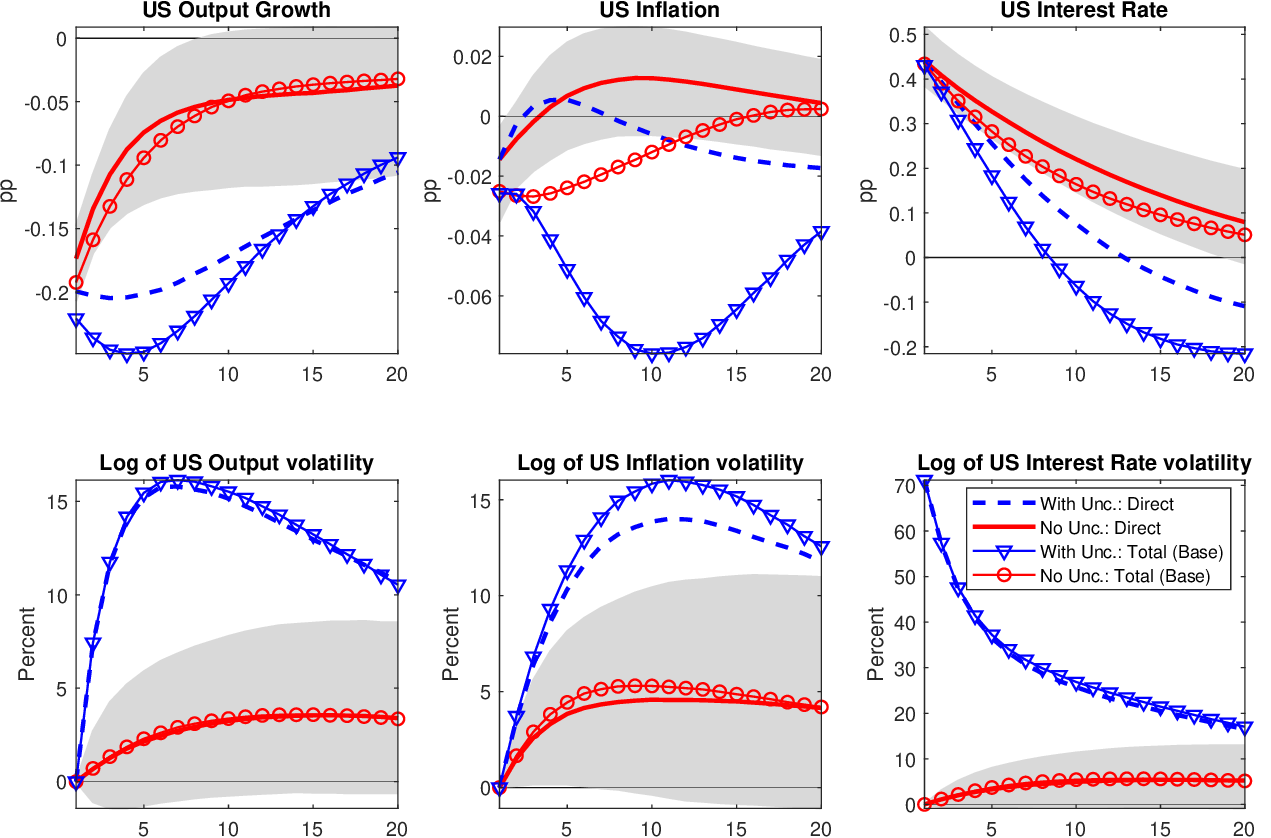}
 \begin{flushleft}
   \footnotesize{Notes: Figure presents the response of US macroeconomy to one standard deviation shock to the US interest rate: in each entry, the red solid line and the shaded area are the median response and the 68 percent intervals where there is \textit{no} shock to US MPU and only direct spillovers are captured; the blue dashed line is the response when one standard deviation shock to the US interest rate occurs with one standard deviation shock to the US policy uncertainty and only direct spillovers are captured. The circle (without uncertainty shock) and triangle (with uncertainty shock) lines are as those in Figure \ref{fig:Impacts-of-USNO-level} taking total spillovers into account.}
 \end{flushleft}
\end{figure}

\begin{figure}[!]
\raggedright{}\caption{\label{fig:Impacts-of-USNO-level-ALL-Direct} Impacts of US Monetary Policy Shocks: Direct versus Total Spillovers (cont)}
\hspace*{-0.5in} \includegraphics[width=0.8\paperwidth,height=0.9\paperwidth]{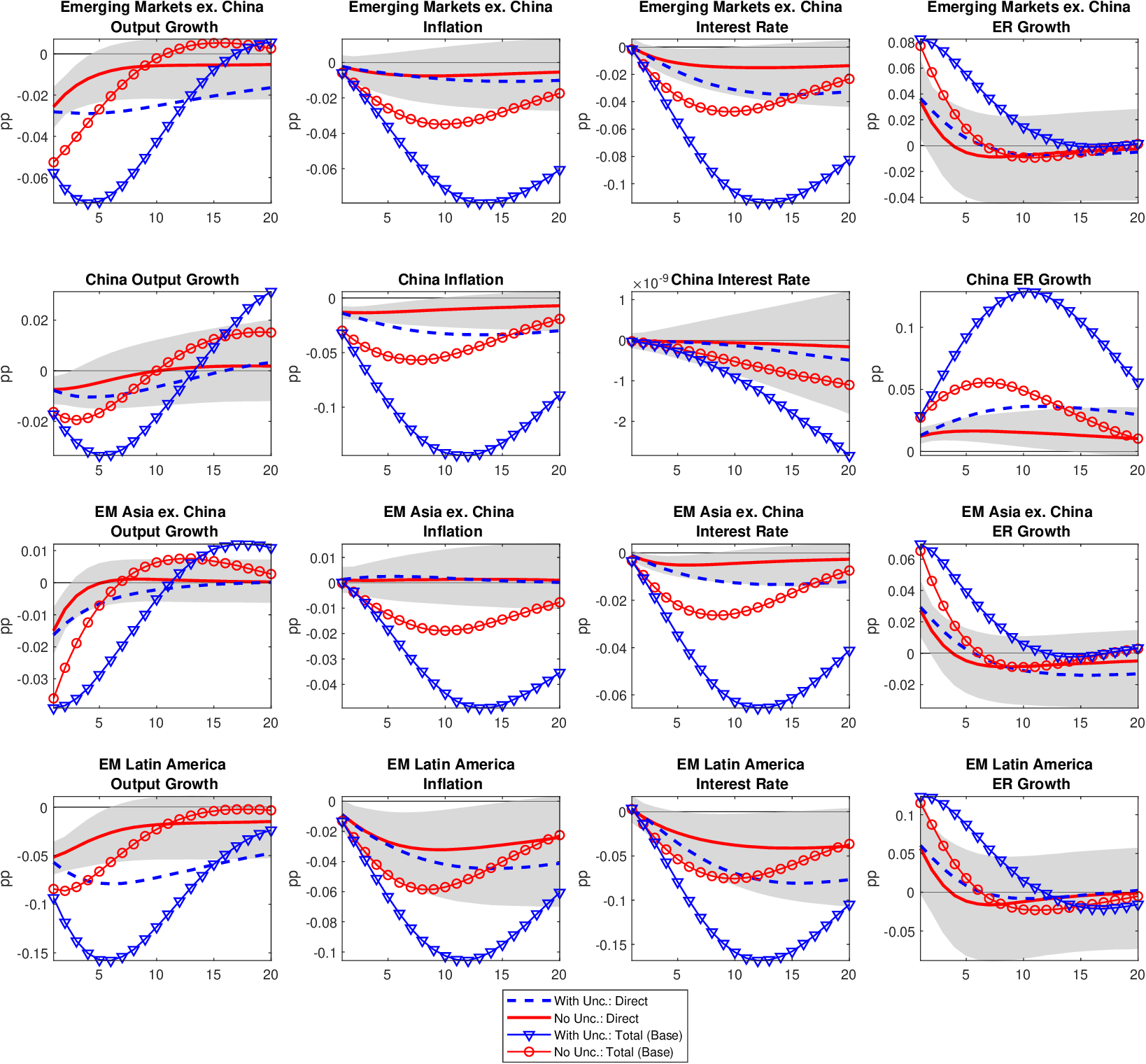}
 \begin{flushleft}
   \footnotesize{Notes: Figure presents the response of emerging markets to one standard deviation shock to the US interest rate: in each entry, the red solid line and the shaded area are the median response and the 68 percent intervals where there is \textit{no} shock to US MPU and only direct spillovers are captured; while the blue dashed line is the response when one standard deviation shock to the US interest rate occurs with one standard deviation shock to the US policy uncertainty and only direct spillovers are captured. The circle (without uncertainty shock) and triangle (with uncertainty shock) lines are as those in Figure \ref{fig:Impacts-of-USNO-level-ALL} taking total spillovers into account.}
 \end{flushleft}
\end{figure}

\subsection{Impacts of US Monetary Policy Shock: Direct versus Total Spillovers}\label{sec:spill}

The next contribution of this paper is visualized in Figure \ref{fig:Impacts-of-USNO-level-US-Direct}, where four cases are covered: direct effects, ignoring changes in other countries due to the US monetary policy changes (line or dashed line), a total effect, where changes in other countries are imported back (spillbacks, denoted by lines with circles and triangles), and cases with (depicted in blue) and without (depicted in red) a positive uncertainty shock. Focusing on standard first-moment shock when keeping the volatility shock shut, the direct effect is substantially reduced when spillback impacts are ignored for inflation. A drop in prices is larger and takes longer when spillback effects are included, as is the case in our estimated GVAR-SV model. The impact on other variables is less affected by the international dimension. However, uncertainty matters big time -- all impacts are significantly amplified when a tightening policy shock is associated with an increase in uncertainty on future policy path. 

The spillbacks also matter -- especially for inflation, but the impacts are larger in magnitude for output growth and interest rate too. We thus provide further empirical support for the literature on global inflation drivers (e.g., \citealp{Auer2019}), global output co-movement (e.g., \citealp{kose2003international}) and global financial cycle (e.g., \citealp{Rey2015,Miranda-Agrippino2020,MirandaAgrippino2020}) in a single empirical framework. Dealing with volatile times when the policymaker faces more uncertainty, and thus the interest rate volatility goes up, leads to larger adjustments, further amplified by the US embedment into the global economy. 

Moving to impacts decomposition for the emerging economies, we document the importance of indirect effects and of policy uncertainty.  When policy tightening is associated with a more uncertain path of future interest rates, the cold caught by the emerging countries is felt  stronger and may require a bit larger dose of medicine (e.g., both Latin America and Asia excluding China tend to lower interest rates more than absent shock to the US monetary policy uncertainty). But the true source of emerging markets catching disease when the US sneezes is the global linkages. Once we factor in how other countries get impacted by the US monetary policy, the effect is considerably amplified since emerging markets are heavily integrated with other economies that also feel the effect of the US monetary policy. 

As exemplified in Figure \ref{fig:Impacts-of-USNO-level-ALL-Direct}, when demand drops in the US, it causes a drop in output, and via the supply chains and final goods trade, in other countries too. Those countries also happen to be trade partners of emerging markets. As a result, emerging markets get impacted directly and indirectly. The most contagious mix, as documented, is when the US tightens monetary policy in uncertain times, requiring substantially larger interest rate movements to counter recessionary and disinflationary total effects as well immense exchange rate depreciation to restore trade balance, raise prices due to more expensive imported goods and services, and increase foreign demand. Inasmuch domestic economy gets back to the growth path, a stronger reaction reduces uncertainty and positively feeds back to the real economy. Of course, these channels are only at work if the country has room for adjusting domestic rates, such as not facing lower bound of interest rates or negative consequences of doing so on exchange rate depreciation and, thus, distress associated with debt denominated in foreign currency.  

To be more precise, a standard deviation tightening shock to the US interest rate, holding other shocks unchanged, leads to around 0.02 percentage point drop in emerging countries (except China) output due the direct effect from the USA (solid red line). When uncertainty is considered, the direct effect becomes \textit{negatively larger}, \textit{longer lasting} and \textit{more persistent}  (blue dashed line). However, the amplification effect stems from US and emerging markets' global integration. Accounting for both direct and indirect effects, EM's output drops by 0.05 percentage points on impact (2.5 times more), but the impact is relatively short-lived and dissipates relatively quicker (circle lines). The story is  different if both -- total spillovers and uncertainty were taken on board -- then output declines even further, with largest drop at around fifth period, and starts dissipating over time (triangle lines). These dynamics are quite robust across variables -- for instance, a direct effect on inflation is bounded by 0.02 percentage points, whereas under a total effect and uncertainty shock, the impact grows fourfold and reaches 0.08 percentage points.  

China's experience has similarities. Its inflation, however, responds most strongly compared other emerging markets. As seen in Figure \ref{fig:Impacts-of-USNO-level-ALL-Direct}, the reason is not in the direct US effect per se, but in Chinese integration into global value chains and imported effects from other countries, coupled with the substantial amplification effect of uncertainty (compare circle and triangle lines). A particularly large impact of uncertainty is shown in the response of renminbi, where the peak occurs after around three years, exceeding the change by three times in the case without a policy uncertainty shock. This result, hence, indicates that China is particularly sensitive to uncertainty fluctuations, with excessively large currency reaction.

Latin America's output drop exceeds Asia's five times when looking at the total effect during uncertain times. For Asia, the direct effect with uncertainty is not distinguishable from the direct effect with no uncertainty, whereas, for Latin America, output drops significantly more when even a direct effect is tracked while allowing for a hike in uncertainty. When it comes to inflation, Asia responds more strongly when global spillovers are included, whereas Latin American countries require both global spillovers and uncertainty shock to obtain a statistically significantly different result. A similar pattern holds for interest rates and exchange rate growth -- Asia's linkages matter even in the case without policy uncertainty shocks, whereas impacts on Latin America depend both on the spillovers (difference between the solid red and the circle red line) and US monetary policy uncertainty (the triangle line). 

\section{\label{sec:Conclusions}Conclusions and Policy Implications}

We documented how US monetary policy affects emerging markets in a multi-country model with endogenous stochastic volatility, with a separation between direct and indirect effects. First, our results show that the Fed's monetary policy tightening reduces the US's output growth and inflation and leads to an increase in US macro uncertainty, especially inflation uncertainty. Once policy tightening is accompanied by an increase in uncertainty on future interest rates path, it leads to deeper recessions. Importantly, we find that a rise in the US interest rate has a negative spillover effect on economic growth in emerging markets. When factoring in policy uncertainty, this tightening effect intensifies, becoming more prominent and longer-lasting. The full impact, considering both direct and indirect effects due to global integration, causes a sharper drop in growth, though it may rebound more quickly. 

Within one internally-consistent global model, we provide empirical support for the literature on global inflation drivers (e.g., \citealp{Auer2019}), global output co-movement (e.g., \citealp{kose2003international}), and global financial cycle (e.g., \citealp{Rey2015,Miranda-Agrippino2020,MirandaAgrippino2020}). Nevertheless, our results highlight substantial heterogeneity in how emerging markets get impacted by comparing the response across China, emerging markets in Asia excluding China, and emerging markets in Latin America. For instance,  China's reaction, especially in inflation, is more pronounced than other emerging markets, and the renminbi's value is particularly susceptible to uncertainty. The demand channel is essential for Latin America, which experiences a fivefold greater drop in output than Asia under total effect during uncertain times. The impact of spillovers and uncertainty on emerging markets are therefore influenced by their exposure to the US, their own trade network, and integration into financial markets, among other factors.

Our results, while estimated with the pre-COVID sample, offer important policy implications for the ongoing rate hikes by the Fed in fighting the inflation surge, which originates from several potential factors, including substantial COVID-related demand support amidst global supply constraints as well as the energy shock from the Russian war in Ukraine, just to name a few. First, it emphasizes the importance of considering the negative spillover to the global economy, which would eventually spillback onto the US economy, in calibrating US monetary policy tightening. Ignoring the spillover and the spillback channel could lead to an over-tightening that reduces growth more than desirable both domestically in the US and globally. This also necessitates the coordination of policy action across countries. 

Second, given that the most critical scenario occurs when the US's monetary policy tightening is accompanied by an increase in policy uncertainty, it emphasizes the need for explicit and precise guidance about the future path of monetary policy. Our study, therefore, finds support for clear communication about the Fed's policy function as well as a commitment to achieve its mandated objectives. 

Third, while the prominent exchange rate depreciation helps dampen the negative spillover impact on output by making emerging market exports cheaper, it could increase the risks of unsustainable debt in emerging countries, particularly for those with a large share of debt in foreign currency. For this latter group, the room to reduce interest rates to stabilize the macroeconomy is limited because doing so could make the exchange rate depreciate even more, and sovereign risks get elevated. At the same time, the accommodative fiscal policy may be outside the scope due to limited fiscal space. For this group, therefore, multilateral support and initiatives to reduce debt distress and foster a quicker return to the path of economic growth are vitally important.

Last but not least, given the important role of US policy in the (global) financial cycle, the possibility of a higher-for-longer rate scenario if inflation becomes harder to tame in the last mile could increase risk to the (global) financial system. Several policy discussions have suggested that other policies could play a certain role in supporting monetary policy in fighting inflation. A prominent candidate is fiscal policy, in which a tighter fiscal stance would allow central banks, including the Fed, to increase interest rates by less than they otherwise would, which would help contain borrowing costs for governments as well as financial stability risks. Such a policy mix also calls for an extension of our analysis to consider fiscal and monetary policy interaction, an interesting and potential avenue that we leave for future research.  

\clearpage{}

\clearpage{}
\newpage\addcontentsline{toc}{section}{References}
\global\long\def\baselinestretch{-1.5}%
\vspace{-1.5in}
 {\small{}{} \bibliographystyle{plainnat}  
\bibliography{Referencesfile_1.bib}
}{\small\par}

\end{document}